\documentclass[final]{aipproc}

\layoutstyle{8x11double}

\newcommand\doingARLO[2][]{%
  \ifx\mmref\undefined #1\else #2\fi
}

\begin{document}

\title{The Effects of Burst Activity on Soft Gamma Repeater Pulse Properties
      and Persistent Emission}


\author{Peter M. Woods}{
  address={Universities Space Research Association \\
  National Space Science and Technology Center \\
  320 Sparkman Dr., Huntsville, AL 35805},
  email={Peter.Woods@nsstc.nasa.gov}
}

\copyrightyear  {2002}

\begin{abstract}

Soft Gamma Repeaters (SGRs) undergo changes in their pulse properties and
persistent emission during episodes of intense burst activity.  SGR~1900$+$14
has undergone large flux increases following recent burst activity.  Both
SGR~1900$+$14 and SGR~1806$-$20 have shown significant changes in their pulse
profile and spin-down rates during the last several years.  The pulse profile
changes are linked with the burst activity whereas the torque variations are
not directly correlated with the bursts.  Here, we review the observed dynamics
of the pulsed and persistent emission of SGR~1900$+$14 and SGR~1806$-$20 during
burst active episodes and discuss what implications these results have for the
burst emission mechanism, the magnetic field dynamics of magnetars, the nature
of the torque variability, and SGRs in general.

\end{abstract}


\maketitle

\section{Introduction}

Soft Gamma Repeaters (SGRs) are an exotic class of high energy transient, very
likely isolated, strongly magnetized neutron stars or ``magnetars.''  For
periods of days to months, SGRs can be found in burst active states where they
emit anywhere from a handful to several hundred bursts.  Typically, the bursts
last $\sim$0.1 sec and have energy spectra (E $>$25 keV) that can be modeled as
a power-law convolved with an exponential.  At lower energies, however, this
empirical model fails to fit the spectrum \cite{olive02}.  The burst energies
follow a power-law number distribution up to $\sim$10$^{42}$ ergs (dN/dE
$\propto$ E$^{-5/3}$ \cite{cheng96,gogus01}), consistent with a so-called
self-organized critical system (e.g.\ earthquakes, Solar flares, etc.
\cite{bak88}) where the burst energy resevoir greatly exceeds the energy output
within any given burst.  On two occasions, more energetic bursts or giant
flares were recorded from SGR~0526$-$66 on 1979 March 5 \cite{mazets79} and
SGR~1900$+$14 on 1998 August 27 \cite{hurley99a,mazets99,feroci99,feroci01}. 
Each of these extraordinary events had a bright ($\sim$10$^{44}$ ergs
s$^{-1}$), spectrally hard initial spike followed by a softer, several minute
long tail showing coherent pulsations at 8 and 5 s, respectively.  More
recently, an intermediate flare ($\sim$10$^{43}$ ergs) lasting 40 s was
recorded from SGR~1900$+$14 on 2001 April 18 \cite{guido01}.

All SGRs are associated with persistent X-ray counterparts; three of them have
quiescent luminosities $\sim$10$^{34}$ ergs s$^{-1}$, while the quiescent flux
level of SGR~1627$-$41 has not yet been determined \cite{kouv02}.  The spectra
of three SGRs can be modeled with a power-law (photon indices $\sim$ 2$-$3.5);
SGR~1900$+$14 requires an additional blackbody component ($kT$ $\sim$0.5 keV
\cite{woods99a,woods01,kouv01}).  Two SGRs show low-amplitude pulsations in
their persistent emission.  The frequency of these pulsations is increasing
rapidly, consistent with the interpretation of an underlying strongly
magnetized neutron star \cite{kouv98}.  For a more comprehensive review of the
properties of SGRs, see \cite{hurley00}.

During the last few years, changes in the X-ray emission properties of SGRs
have been noted during episodes of burst activity \cite{woods01,ibrahim01}. 
Through studying the transient effects imparted upon SGRs (or the lack thereof)
during times of burst activity, we have gained deeper insight into the nature
of the burst mechanism and the SGR systems in general.  Here, we review the
observed influence of burst activity on SGR pulse properties and persistent
X-ray emission, limiting ourselves to the two SGRs that show pulsations in
their X-ray emission, namely SGR~1900$+$14 and SGR~1806$-$20.

\begin{figure}

\caption{{\it Top panel} -- Burst rate history of SGR~1900$+$14 as observed
with BATSE.  {\it Middle panel} -- Persistent/Pulsed flux history of
SGR~1900$+$14 covering 4.5 years.  The left vertical scale is unabsorbed 2$-$10
keV flux and the right is pulsed flux in units of counts s$^{-1}$ PCU3$^{-1}$. 
The dotted line marks the nominal quiescent flux level of this SGR.  Note that
the spike in the pulsed flux shortly after MJD 52000 coincides with a burst
active episode not covered by the BATSE monitoring ({\it top}).  See text for
further details.  {\it Bottom panel} -- Pulse fraction of SGR~1900$+$14 (2$-$10
keV) as measured within the four BeppoSAX observations using the MECS
instruments.  The dashed line marks the mean RMS pulsed fraction ($f_{\rm RMS}
\sim$ 0.11).}

\includegraphics[width=.9\textwidth]{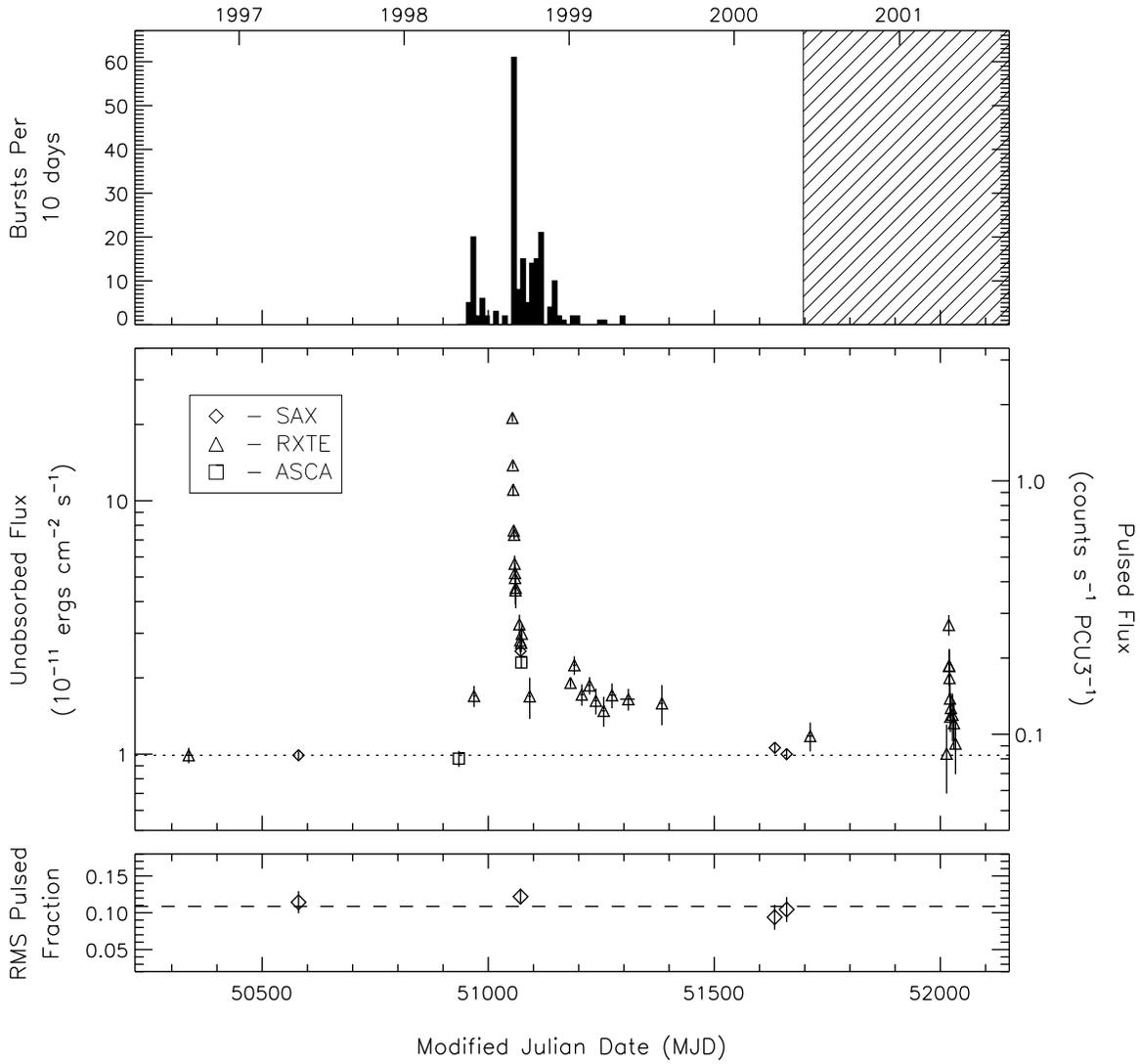}
\end{figure}

\section{Persistent and Pulsed Flux}

Changes in the flux of SGRs was first noted in SGR~1900$+$14 following the
giant flare of August 27 \cite{remillard98,kouv99,murakami99,woods99a}. 
Following this discovery, a compilation of persistent and pulsed flux
measurements over several years (Figure 1 \cite{woods01}) revealed that, in
general, there is an excellent correlation between burst activity (top) and
enhancements in the persistent/pulsed flux from this SGR (middle).  We have
found that the pulse fraction (bottom) is consistent with remaining constant at
most epochs despite changes in the persistent flux.  It is by assuming that
this fraction remains constant at all times that we can plot both the pulsed
flux ({\it RXTE} PCA) and the persistent flux ({\it BeppoSAX} and {\it ASCA})
on the same scale.  We note, however, that there are exceptions to this rule
when the pulse fraction has increased for short periods of time (see below).

We have found that the brightest pulsed/persistent flux excess seen in Figure 1
is directly linked with the August 27 flare.  The excess decays approximately
as a power-law in time ($F \propto t^{-0.7}$) following the giant flare (Figure
2 \cite{woods01}), qualitatively similar to GRB afterglows.  In order to avoid
confusion between the two phenomena, we will refer to the excesses in SGRs as
X-ray tails rather than afterglows hereafter.  The spectrum (0.1$-$10 keV) of
the X-ray tail at $\sim$19 days after the flare was found to be exclusively
non-thermal \cite{woods99a}.

\begin{figure}[!htb]

\caption{The flux decay following the 1998 August 27 flare from SGR~1900$+$14. 
The reference time is the beginning of the flare as observed in soft
$\gamma$-rays.  The dotted line is a fit to the RXTE/PCA, BeppoSAX, and ASCA
data only (i.e.\ the ASM data are not included in the fit).  The slope of this
line is $-$0.713 $\pm$ 0.025.}

\includegraphics[width=.45\textwidth]{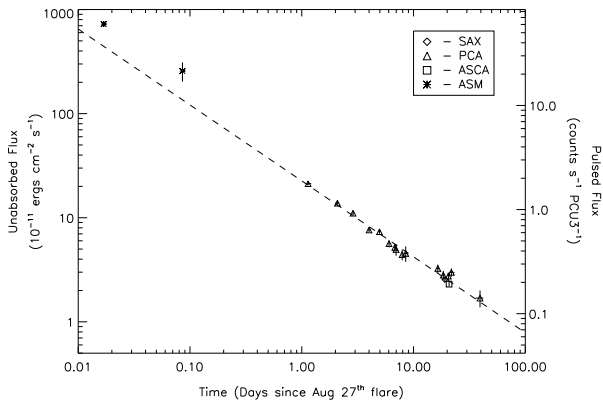}
\end{figure}

For SGR~1900$+$14, there are now four X-ray tails that can be linked with
specific bursts or flares.  The second of these events was recorded on 1998
August 29 (Figure 3 \cite{ibrahim01}).  This burst had a high gamma-ray fluence
and an X-ray tail whose bolometric flux decayed approximately as a power-law in
time.  The spectrum of this tail softens with time.  Formally, the spectrum is
equally well fit by a power-law plus a blackbody or a thermal bremsstrahlung,
each with interstellar attenuation \cite{ibrahim01}.  However, the
bremsstralung model yields a column density $\sim$5 times larger than the
measured column from the persistent emission whereas the two component model
fit yields a column consistent with the persistent emission value.  The pulsed
fraction increases above the quiescent level (11\% RMS) up to $\sim$20\% during
this tail (Figure 4 \cite{lenters02}), and the phase of the pulsations do not
shift during the tail relative to the pre-burst pulse phase.

\begin{figure}[!htb]

\caption{The energetic burst of 1998 August 29 from SGR~1900$+$14 as seen with
BATSE (top panel) and the RXTE PCA (bottom panel).}

\includegraphics[width=.45\textwidth]{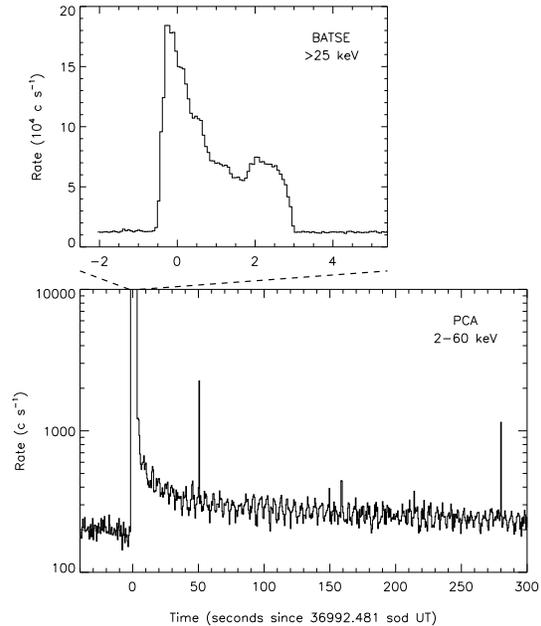}
\end{figure}

The last two bursts with X-ray tails were detected on 2001 April 18 and April
28.  Spectral analysis of the April 18 burst tail is presented in
\cite{kouv01,feroci02,fox02}.  The April 28 event is discussed in greater
detail elsewhere in this volume \cite{lenters02}.  During each of these events,
the pulse fraction was found to increase during the tail
\cite{woods02b,lenters02}.  The spectrum of the April 28 tail is a cooling
blackbody \cite{lenters02}, dissimilar to the 1998 August tails which each
required a power-law component.

\begin{figure}[!htb]

\caption{The evolution of the 2$-$10 keV pulse fraction during the X-ray tail
following the burst of 1998 August 29.  The dotted line denotes the average
pulse fraction observed during quiescence.}

\includegraphics[width=.45\textwidth]{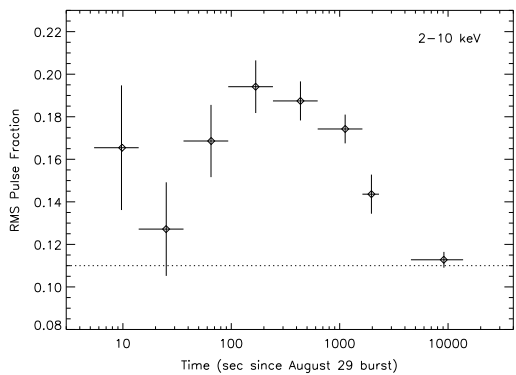}
\end{figure}

Even though we have detected just four X-ray tails following energetic bursts
from SGR~1900$+$14, we find significant differences between them.  First, there
are varying levels of thermal and non-thermal emission within the tails.  Also,
the pulse fraction increases by up to a factor $\sim$3 in one case (April 28)
and not at all in another (August 27).  An interesting trend which arises from
this small set of X-ray tails is that the pulse fraction enhancement in the
separate X-ray tails appears to correlate with the magnitude of the thermal
contribution to the X-ray flux.  That is, tails with the highest relative
blackbody flux show the largest increase in pulse fraction.

\section{Pulse Profiles}

Currently, the pulse profiles of both SGR~1900$+$14 and SGR~1806$-$20 are very
nearly sinusoidal (i.e.\ they show very little power at the higher harmonics). 
This has not always been the case, however, as both SGRs have shown significant
changes in their pulse profiles during the last several years.  The most
notable of which was the dramatic change in the pulse profile of SGR~1900$+$14
during the tail of the giant flare of August 27
\cite{hurley99a,mazets99,feroci99,feroci01}.

Forty seconds after the onset of the August 27 flare, 5.16 s coherent gamma-ray
pulsations at high amplitude emerged.  Initially, the pulse profile was
complex, having four distinct maxima per rotation cycle.  Toward the end of the
flare, the pulse profile was significantly more sinusoidal (Figure 5 -- middle
row).  The same qualitative behavior was observed in the persistent X-ray
emission from SGR~1900$+$14.  In all observations prior to 1998 August 27, the
pulse profile was complex having significant power at higher harmonics (Figure
5 -- top row).  For all observations after August 27 through early 2000, the
pulse profile remained relatively simple (Figure 5 -- bottom row).  Hence, the
pulse profile change observed at gamma-ray energies during the tail of the
August 27 flare translated to the persistent emission from this SGR in a
sustained manner (i.e.\ for years after the August 27 X-ray tail had
disappeared) \cite{woods01}.

\begin{figure}

\caption{Evolution of the pulse profile of SGR~1900$+$14 covering 3.8 years. 
All panels display two pulse cycles and the vertical axes are count rates with
arbitrary units.  The two middle panels were selected from Ulysses data
(25$-$150 keV) of the August 27$^{\rm th}$ flare.  Times over which the Ulysses
data were folded are given relative to the onset of the flare (T$_{\rm o}$). 
See text for further details.  The top and bottom rows are integrated over the
energy range 2$-$10 keV.  From top-to-bottom, left-to-right, the data were
recorded with the RXTE, BeppoSAX, ASCA, RXTE, RXTE, RXTE, BeppoSAX, and RXTE.}

\includegraphics[height=.9\textheight]{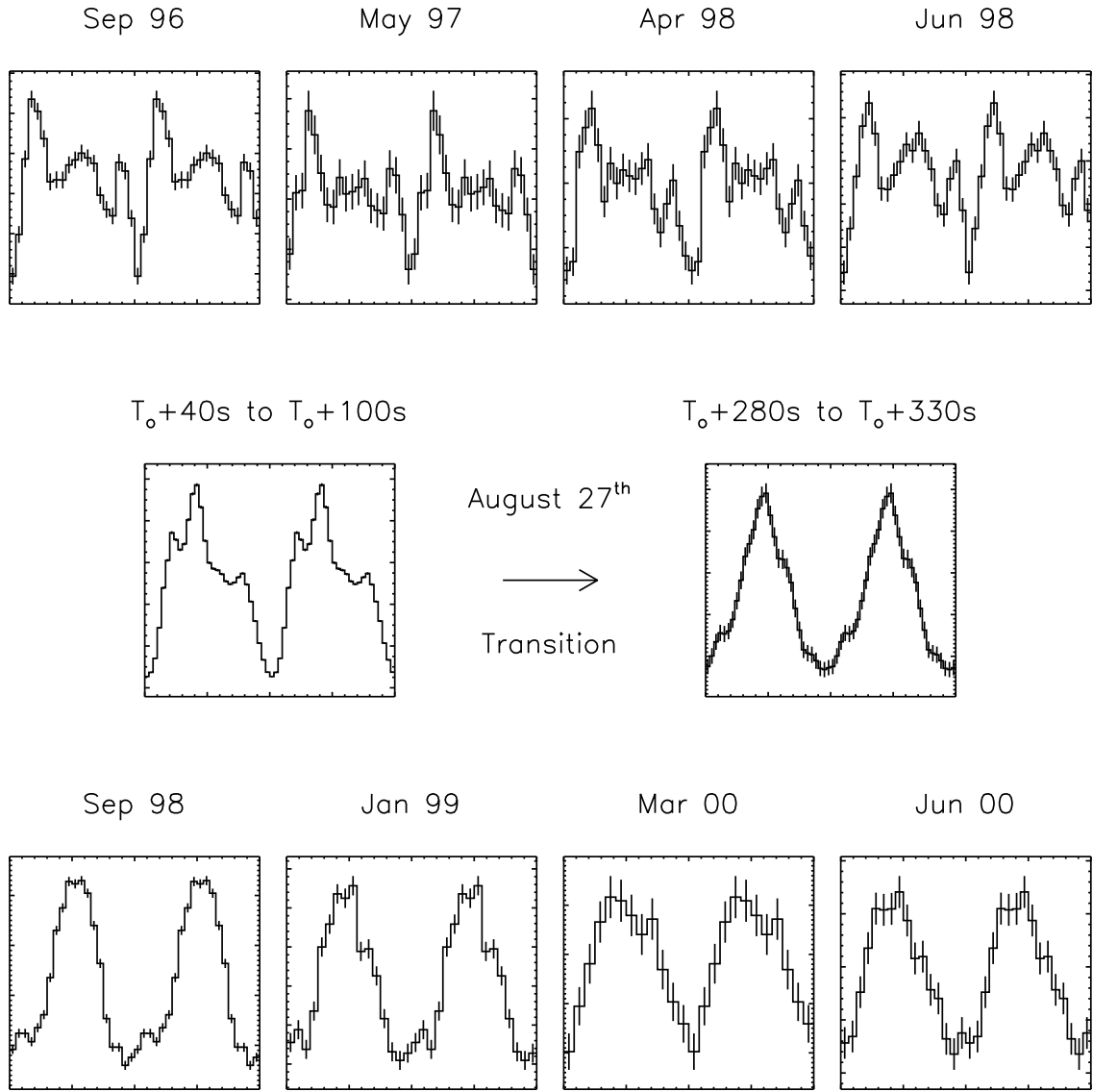}
\end{figure}

By the middle of the year 2000, the pulse profile began to show slightly more
power in the higher harmonics \cite{gogus02}.  The next observations of
SGR~1900$+$14 took place in the hours and days following the April 18
intermediate flare.  At some point between the latter half of 2000 and the days
directly after the April 18 flare, the pulse profile simplified in shape
\cite{woods02b}.  As with the August 27 flare, the direction of the pulse
profile change here was the same in that power at the higher harmonics lessened
following the flare \cite{gogus02}.

A systematic study of the temporal and spectral evolution of the pulse profile
of SGR~1900$+$14 using exclusively {\it RXTE} PCA observations has recently
been completed \cite{gogus02}.  In this study, we show that there is a
significant energy dependence of the pulse profile, particularly when the
profile was complex in shape (i.e.\ prior to 1998 August 27).  Moreover, the
pulsed flux spectrum during the X-ray tail of the August 27 flare becomes
harder with time.

The evolution of the pulse profile of SGR~1806$-$20 is also presented in
\cite{gogus02}.  We find significant temporal evolution in the pulse profile of
this SGR from 1996 November to 1999 January.  Due to the sparseness of the
observations, however, we cannot determine the exact time of this change, nor
the timescale over which it progressed to better than 2.3 years.

\section{Pulse Timing}

Coherent pulsations from the persistent emission of SGR~1806$-$20 were
discovered within an {\it RXTE} PCA observation from 1996 November
\cite{kouv98}.  From archival observations, it was found that the spin
frequency of this SGR was decreasing rapidly, indicative of a strongly
magnetized neutron star spinning down via magnetic braking \cite{kouv98}.  The
spin frequency history of this SGR now extends from 1993 through 2001 (Figure 6
\cite{woods02a}).  We have found that at all times, the SGR has been spinning
down, but the rate of spindown shows substantial variability.  In fact, the
measured spin-down torque on this SGR has been found to vary by up to a factor
$\sim$4.  Unlike the flux variability of SGR~1900$+$14, the torque variations
seen in SGR~1806$-$20 do not correlate with the burst activity \cite{woods02a}.

\begin{figure}

\caption{{\it Top} -- Burst rate history of SGR~1806$-$20 as observed with
BATSE.  The hashed region starts at the end of the {\it CGRO} mission.  {\it
Middle} -- The frequency history of SGR~1806$-$20 covering 7.1 years.  Plotting
symbols mark individual frequency measurements and solid lines denote
phase-connected timing solutions.  The dashed line marks the average spin-down
rate prior to burst activation in 1998.  {\it Bottom} -- The frequency
derivative history over the same timespan.  Dotted lines denote average
frequency derivative levels between widely spaced frequency measurements. 
Solid lines mark phase-coherent timing solutions and triangles mark
instantaneous torque measurements, both using {\it RXTE} PCA data.}

\includegraphics[height=.75\textheight]{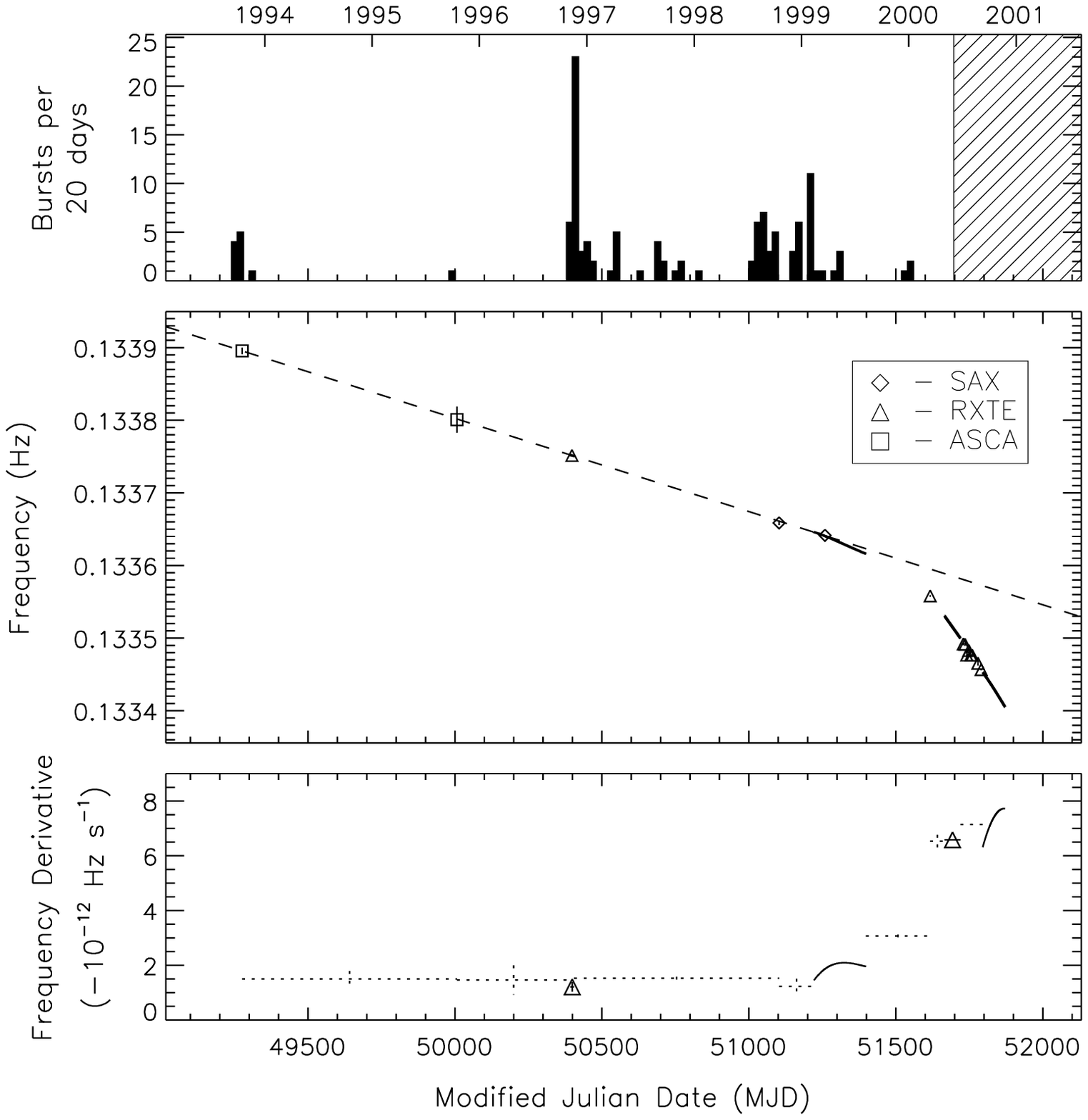}
\end{figure}

Pulsations from the X-ray counterpart of SGR~1900$+$14 were discovered during
an ASCA observation in 1998 April \cite{hurley99b}, shortly before the SGR
entered an intense, sustained burst active interval \cite{hurley99c}.  Similar
to SGR~1806$-$20, subsequent observations showed that this SGR was spinning
down rapidly and irregularly \cite{kouv99,woods99b}.  The spin frequency
history of this SGR now extends from 1996 through 2001 (Figure 7
\cite{woods02a}).  As with SGR~1806$-$20, the variations in torque do not
directly correlate with the burst activity from this SGR with one notable
exception, the giant flare of August 27.

\begin{figure}

\caption{{\it Top} -- Burst rate history of SGR~1900$+$14 as observed with
BATSE.  The hashed region starts at the end of the {\it CGRO} mission.  {\it
Middle} -- The frequency history of SGR~1900$+$14 covering 4.7 years.  Plotting
symbols mark individual frequency measurements and solid lines denote
phase-connected timing solutions.  The dashed line marks the average spin-down
rate prior to burst activation in 1998.  {\it Bottom} -- The frequency
derivative history over the same timespan.  Dotted lines denote average
frequency derivative levels between widely spaced frequency measurements. 
Solid lines mark phase-coherent timing solutions and triangles mark
instantaneous torque measurements, both using {\it RXTE} PCA data.}

\includegraphics[height=.75\textheight]{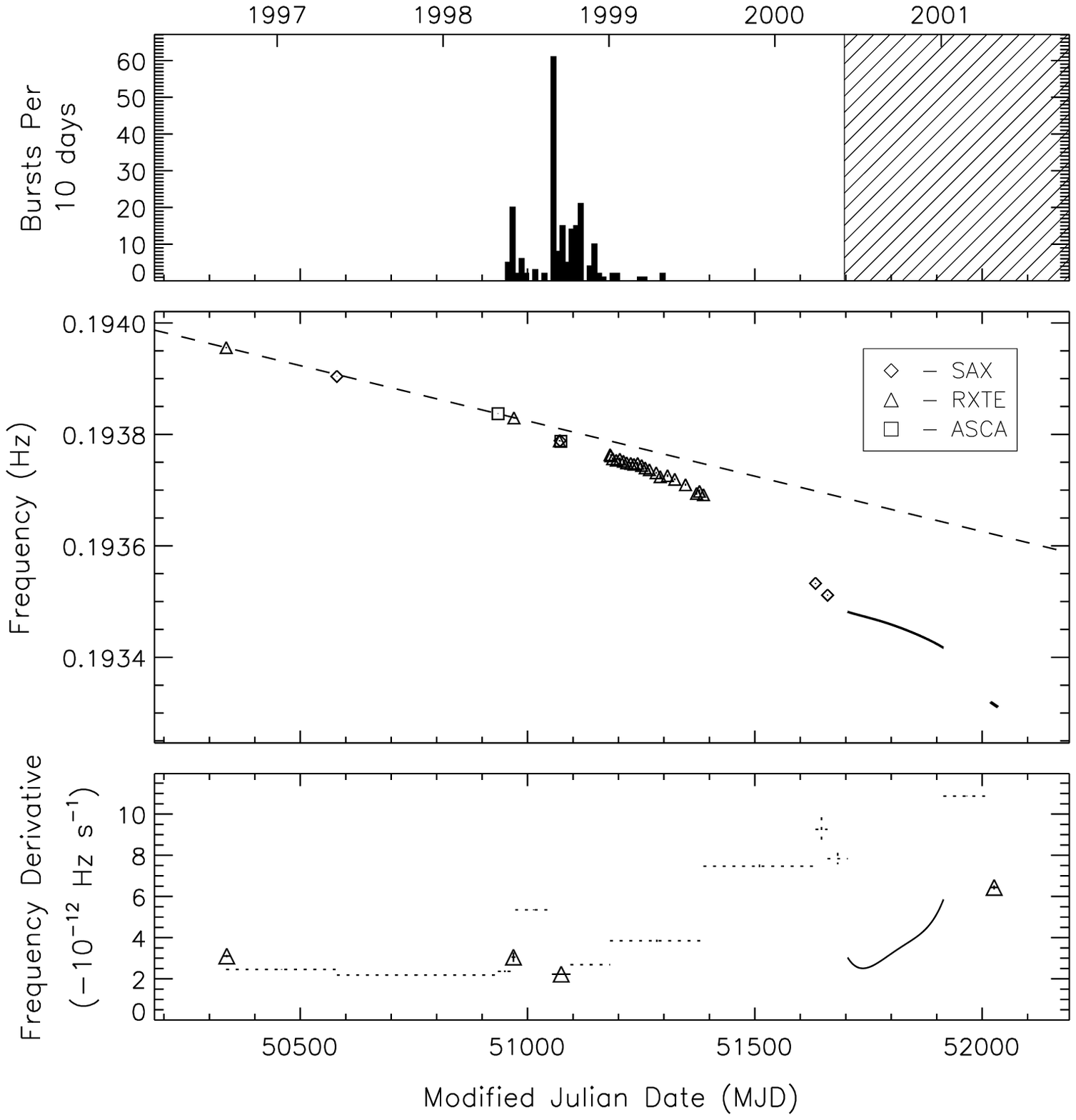}
\end{figure}

From an earlier compilation of pulse frequency measurements between 1996
September and 1999 January, we showed that the spin down of this SGR showed
small variations, yet remained constant on average for timescales longer than
about a month at nearly all epochs \cite{woods99b}.  The lone exception to this
rule was an 80 day interval during the middle of 1998 where the average
spin-down rate nearly doubled.  Contained within this 80 day interval was the
August 27 flare and we argued that this spin-down anomaly was likely linked to
the flare \cite{woods99b}.  Subsequently, Palmer \cite{palmer01} showed that
the phase of the pulsations in gamma-rays was offset from a backward
extrapolation of the X-ray pulse train recorded during the days following the
flare, supporting our earlier inference that the star had spun down rapidly
during/after the giant flare.  The conclusion of rapid spin down depends
critically on the energy dependence of the pulse profile.  More recent
observations of gamma-ray pulsations during the 2001 April 18 flare have shown
that there is very little change in the pulse profile with energy for this
source \cite{woods02b}, substantiating the claim that the phase offset in the
August 27 flare was due to a sudden change in torque, perhaps from a
relativistic particle outflow from the stellar surface.  Gamma-ray observations
during the flare \cite{feroci01} and a transient radio nebula discovered
following the flare \cite{frail99} provide independent evidence for the
existence of a particle outflow.

As illuded to in the previous paragraph, we searched for a similar effect in
the aftermath of the April 18 intermediate flare.  We found that unlike the
August 27 flare, the phase of the gamma-ray pulsations matched nicely with the
X-ray ephemeris \cite{woods02b}.  A fortuitous X-ray monitoring observation 4
days prior to the flare shows that there was a short timescale change in the
spin ephemeris somewhere between April 14 and April 18, however, the brevity of
the April 14 observation precluded us from constraining the manner in which
this change occurred.

Finally, we note that although the August 27 flare likely did alter the spin
down of SGR~1900$+$14, its impact was very small relative to the much larger
variations observed during $\sim$5 years of monitoring.  So, in general, the
direct effects of burst activity are insignificant to the overall torque noise
in each of these SGRs.  For a more complete discussion of the torque
variability in these two SGRs, as well as a quantitative analysis of the torque
noise, see \cite{woods02a}.

\section{Discussion}

We have summarized the recent observations of dynamic behavior in the
persistent and pulsed emission from SGR~1900$+$14 and SGR~1806$-$20.  Now, we
will discuss what constraints these observations place on the models for the
SGRs, in particular the magnetar model.

The magnetar model postulates that the SGRs are young neutron stars with
super-strong magnetic fields ($B \sim 10^{14}-10^{15}$ G).  It is the decay of
this strong field which powers both the burst and persistent emission
\cite{td95,td96}.  The steady X-ray emission is generated by persistent
magnetospheric currents and low-level seismic activity within and beneath the
stellar surface.  The burst emission is due to the build up of stress in the
stellar crust from the evolving magnetic field and the eventual release of this
stress when the crust fractures.  To date, this model provides the most
accurate description of the persistent, pulsed, and burst properties of SGRs.

Currently, only four clear X-ray tails have been detected, all from
SGR~1900$+$14.  As mentioned earlier, there is a potential correlation between
the relative abundance of thermal emission in the tail and the enhancement of
the pulse fraction.  This correlation, if proved correct with the detection and
analysis of several more SGR tails, would provide a strong argument for heating
of a localized region on the neutron star during bursts.  Since the pulse
fraction increases during some of these tails, the flux enhancement must be
anisotropic about the star.  In the cases of the August 29 and April 28 burst
tails, the location of the heating is also constrained.  In each of these
events, we have precise pulse phase information prior to, during, and after the
tail.  For both bursts, the phase of the pulsations during the tail does not
shift relative to the pulse phase prior to the burst.  This requires that the
localized region on the neutron star with the largest relative flux enhancement
is the same region giving rise to the persistent X-ray pulse peak (e.g.\ the
polar cap).

With regards to the magnetar model, localized heating of the polar cap requires
that the fracture site of the burst be at the same location.  The peak of the
August 29 flare in gamma-rays lags behind the centroid of the pulse profile
peak in X-rays by $\sim$0.1 cycle, although the burst light curve (duration
$\sim$3.5 s) covers a large fraction of a pulse cycle spanning the pulse
valley.  The phase alignment of the April 28 burst is yet to be determined.  
Measuring the phase alignment of the April 28 burst and detailed modeling of
the expected lag between the peak in the X-ray pulse profile and the rise
and/or peak of the burst are required before one can determine whether or not
these two bursts fit with the picture of a localized fracture region near the
polar cap. 

The dramatic change in the pulse profile of SGR~1900$+$14 in conjunction with
the giant flare requires a substantial change in the magnetic field of the
neutron star \cite{woods01,thompson00}.  In the magnetar model, there are at
least two possible ways this can happen.  One possibility is that a twist in
the magnetosphere is generated following the flare, driving a persistent
current which produces an optically thick scattering screen at some substantial
distance ($\sim$10 $R_*$) from the stellar surface.  In this model, the surface
field geometry remains complex at all times.  The pulse profile, however,
simplifies when the scattering screen is present (i.e.\ after the flare).  The
scattering screen must have the properties of redistributing the radiation in
phase, but not in energy in order to account for the reemergence of the
blackbody component after the August 27 tail fades away \cite{tlk02}.  The
decay of this magnetospheric twist is believed to be several years.  An
alternative scenario involves restructuring of the surface magnetic field
geometry.  In this picture, the field geometry is complex prior to the flare
and relaxes to a more dipolar structure following the event giving rise to the
observed change in pulse profile.

Thompson, Lyutikov \& Kulkarni \cite{tlk02} recently investigated each of these
scenarios in detail, noting advantages and disadvantages for each model.  In
this work, they have identified further observational tests involving the
energy spectrum of the emission before and after the flare.  Simulations of the
expected behavior \cite{thompson02} and an analysis of the spectral evolution
of SGR~1900$+$14 \cite{woods02c} are currently underway to work towards
resolving this issue.

Unlike the flux variability, the torque enhancements in these systems do not
correlate with the burst activity.  In the context of the magnetar model, the
abscence of a direct correlation between these two parameters has strong
implications for the underlying physics behind each phenomenon.  The magnetar
model postulates that the bursting activity in SGRs is a result of fracturing
of the outer crust of a highly magnetized neutron star.  Furthermore, the
majority of models proposed to explain the torque variability in magnetars
invoke crustal motion and/or low-level seismic activity
\cite{tb98,hck99,thompson00}.  Since there is no direct correlation between the
burst activity and torque variability, then either ($i$) the seismic activities
leading to each observable are decoupled from one another, or ($ii$) at least
one of these phenomena is {\it not} related to seismic activity
\cite{woods02a}.  Simultaneous spectral information from imaging X-ray
telescopes (e.g.\ {\it BeppoSAX}, {\it Chandra}, and {\it XMM-Newton})
complimentary to the torque measurements obtained with the {\it RXTE} PCA would
be useful in determining the nature of the torque variabilitity in these SGRs.

\begin{theacknowledgments}

I thank the many collaborators who have contributed to the results discussed
here.  I would like to thank Chryssa Kouveliotou for many useful discussions
and a careful reading of the manuscript.  I acknowledge my own support from the
Long Term Space Astrophysics program (NAG 5-9350).

\end{theacknowledgments}

\newpage


\doingARLO[\bibliographystyle{aipproc}]
          {\ifthenelse{\equal{\AIPcitestyleselect}{num}}
             {\bibliographystyle{arlonum}}
             {\bibliographystyle{arlobib}}
          }
\bibliography{burst_influence}

\end{document}